\documentclass[reprint,superscriptaddress,amsmath,amssymb,aip,jap,floatfix]{revtex4-1}
\usepackage[version=3]{mhchem}
\usepackage{siunitx,natmove,graphicx}

\def\df{d_\text{f}}
\def\Ef{E_\text{f}}
\def\Efeff{E_\text{f}^\text{eff}}
\def\Cf{C_\text{f}}
\def\Ctot{C_\text{tot}}
\def\Csys{C_\text{sys}}
\def\sigf{\sigma_\text{f}}

\DeclareSIUnit{\um}{\micro\m}

\begin{document}

\def\SINANO{Suzhou Institute of Nano-Tech and Nano-Bionics, Chinese Academy of Sciences, Suzhou 215123, China}
\def\SHUM{Department of Polymer Materials, School of Materials Science and Engineering, Shanghai University, Shanghai 200444, China}
\def\SHUC{Department of Chemistry, College of Sciences, Shanghai University, Shanghai 200444, China}

\title{Evaluating the System Compliance in Testing Carbon Nanotube Fibers}

\author{Chunyang Jiang} \affiliation{\SINANO} \affiliation{\SHUM}
\author{Xin Zhang} \affiliation{\SINANO} \affiliation{\SHUC}
\author{Jingna Zhao}
\author{Jingyun Zou}
\author{Xiaohua Zhang}
\email{zhangxhcm@gmail.com}
\affiliation{\SINANO}


\begin{abstract}
System compliances of two testing machines (Keysight T150 and Instron 3365) were studied for the tensile testing of
carbon nanotube (CNT) fibers. Due to the different gripping methods, the T150's system compliance was nearly zero, while
that of the Instron was nonnegligible. By analyzing the data dispersion with the Weibull theory, a safe gauge length of
$>$10 mm was suggested for CNT fibers. Therefore, a standard test procedure should involve the compliance evaluation,
choice of gauge length, and Weibull statistical analysis.
\end{abstract}

\maketitle

\section{Introduction}

Fibers are the principal load carrying component in their reinforced composites. Thus it is important to accurately
measure the fiber strength, modulus, strain at break, and toughness. However, much care is needed for the testing of
different types of fiber. For example, for an ultrafine fiber with a diameter smaller than several \si{\um}, the fiber
specimen should be very carefully gripped in order to minimize any undue stress, and a correction for system compliance
is also very essential for the determination of Young's modulus \cite{li1985improvement}. For a brittle specimen, a
template-gripping procedure rather than the direct gripping is suggested to avoid premature failure initiation sites
\cite{ac1557, kant.m:2014}. Furthermore, most material test systems, i.e., Instron and Keysight testing machines,
measure crosshead or actuator displacement that can be used as a measurement of specimen deformation. However, the
displacement output recorded by the system is actually the sum of the system compliance and the specimen deformation.
Thus, the introduction of a template, for example a paper card, could increase the system compliance due to the grip
deformation. 

On the other hand, although various test standards are available for synthetic fibers \cite{ac1557, ad1557, ad3822,
ad2256, GBT31290}, there is still lack of direct application of these standards on the newly-developed carbon nanotube
(CNT) fibers. As a type of assembly material \cite{zhao.jn:2015}, the tensile strength and Young's modulus of CNT fiber
strongly depend on the densification level \cite{li.s:2012}, twisting level \cite{zhao.jn:2010}, and the tube structure
\cite{jia.jj:2011}. Even for a given CNT fiber, its mechanical properties were found to vary remarkably upon the strain
rates; when the strain rate changed from $2\times10^{-5}$ to $2\times10^{-1}$ \si{\per\s}, there could be $\sim$100\%
increase in strength for a loosely packed CNT fiber \cite{zhang.yn:2012}. Obviously, a new test standard should be
proposed for such assembly fibers. Furthermore, due to the lack of test standard, most tensile tests on CNT fibers were
performed with a gauge length smaller than or close to 10 mm \cite{zhang.m_2004, zhong.xh_2010, meng.fc:2014,
ryu.s_2015}. Extremely, in several studies a very short length of 1--2 mm was used \cite{koziol.k:2007, wang.jn:2014}.

Recently, Behabtu \textit{et al.} \cite{behabtu.n:2013} suggested a gauge length of 20 mm to test CNT fibers, in
agreement with a Chinese Standard which suggests a length of 25 mm for single carbon fibers \cite{GBT31290}. For CNT
fibers, a short gauge length can characterize the intrinsic strength of the CNT bundles that constitute the fiber,
whereas macroscopic gauge lengths measure the extrinsic properties of the fiber, due to the cohesion between CNT bundles
in the network \cite{vilatela.jj:2011}. Therefore, towards a suitable test standard for CNT fiber, it is important to
pay more heed to the testing technique itself.

Here, we remedy such aspect by means of system compliance evaluation. Three types of CNT fibers and a carbon fiber were
tested by two different testing machines. Due to the different gripping methods, the Instron testing machine that we
used exhibited a nonnegligible system compliance while the compliance for the Keysight machine was nearly zero. For the
latter, although the compliance value was very small, it generally decreased as the fiber's modulus increased. The
compliance evaluation also clearly showed that the spread of both the modulus and strength became smaller with
increasing the gauge length. By further considering the Weibull distribution, a standard test procedure was provided for
CNT fibers, including the compliance evaluation, choice of gauge length, and statistical strength analysis.

\section{Experimental}

\subsection{Materials}

\begin{table*}[!t]
\caption{Fiber diameter, Young's modulus, tensile strength, failure force, and strain at break for the four types of
fiber. The values were obtained from measurement except that the carbon fiber's diameter was input as a constant
according to the product's data sheet \cite{t700sc}.}
\label{tab:fiber}
\begin{tabular}{cccccc}
\hline\hline
Fiber & Diameter (\si{\um}) & Modulus (GPa) & Strength (MPa) & Failure force (mN) & Strain at break (\%) \\
\hline
CNTF-A & 13.9--24.6 & 8.1--24.5 & 294--911 & 53.3--153.7 & 2.9--7.8 \\
CNTF-B & 13.9--21.3 & 12.1--30.7 & 250--718 & 61.4--177.2 & 1.9--4.4 \\
CNTF-C & 10.0--13.3 & 57.0--91.5 & 917--1779 & 77.8--247.3 & 2.0--3.2 \\
CF-T700SC & 7 & 198--261 & 2722--5743 & 104.7--221.0 & 1.1--2.5 \\
\hline\hline
\end{tabular}
\end{table*}

CNT fibers and carbon fibers were tested in this study. For a better comparison, these fibers had a large variation in
modulus. The CNT fibers were obtained by a forest-based spinning method \cite{zhao.jn:2010, jia.jj:2011, li.s:2012},
where a CNT sheet was pulled out from a vertically aligned CNT forest and fabricated under drawing and twisting into a
continuous fiber. The details of the fiber spinning have been reported in our previous studies \cite{meng.fc:2014}. By
using ethylene glycol (EG) densification \cite{li.s:2012, meng.fc:2014}, a high modulus was achieved for the CNT fibers,
up to $\sim$90 GPa, while without such treatment the modulus was just up to $\sim$30 GPa. In the present study, three
CNT fibers were tested, namely CNTF-A, CNTF-B, and CNTF-C respectively. The CNTF-A and CNTF-B were not densified with
solvent while the CNTF-C was densified with EG. Toray T700SC-12K-50C carbon fibers were used as a comparison to CNT
fibers \cite{t700sc}. The untreated carbon fibers were desized by acetone inside a sealed flask for 12 h, and then dried
naturally. Table \ref{tab:fiber} lists the fiber diameter ($\df$), Young's modulus ($\Ef$), tensile strength ($\sigf$),
and failure strain for these fibers.

\subsection{Tensile tests}

\begin{figure}[!t]
\centering
\includegraphics[width=.36\textwidth]{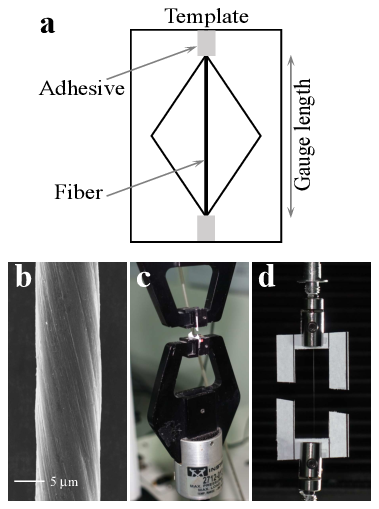}
\caption{(a) Paper template used to mount fiber sample. (b) A scanning electron microscopy image of an EG-densified CNT
fiber. (c,d) Two fiber samples mounted by the Instron 2712-013 grips and T150 split-shim grips. The gauge lengths were 6
and 24 mm respectively. In order to make the fiber more visible by eyes, the two fibers mounted here both had a diameter
over 50 \si{\um}.}
\label{fig:tab}
\end{figure}

By following Kim \textit{et al.} \cite{kim2013effects, kim2015effect} and a Keysight Application Note
\cite{keysight9505}, the 210 \si{\g\per\square\meter} photo paper was used as the template whose shape was schematically
shown in Figure \ref{fig:tab}a. A typical fiber morphology was shown in Figure \ref{fig:tab}b. As compared to the 90
\si{\g\per\square\meter} paper \cite{fidelis2013effect}, the present choice could benefit reducing more of the system
compliance. Cyanoacrylate glue (commonly known as instant adhesive) was used to mount the fibers onto the paper's
printing surface.

Two testing machines, a T150 Universal Testing Machine (Keysight Technologies, Inc., Santa Rosa, USA) and an Instron
3365 Universal Testing Machine (Instron Corp., Norwood, USA), were used to perform the compliance study. The
specifications of the T150 include a maximum load of 500 mN, load resolution of 50 nN, displacement resolution $<$0.1
nm, and extension resolution of 35 nm. For the Instron 3365 equipped with a 10-N load cell, the load resolution is 0.5
mN and the minimum displacement rate is 0.01 \si{\milli\meter\per\minute}. To avoid the effect of strain rate
\cite{zhang.yn:2012}, a fixed strain rate of $2\times 10^{-4}$ \si{\per\s} was used for all tests. Four different gauge
lengths were used: 6, 12, 24, and 40 mm. All testing was conducted in a quasistatic mode at ambient temperature
($\sim$25 \si{\celsius}) and a relative humidity of about 40\%. The diameter was measured by the customary optical
diffraction model, where a 532-nm green laser was used \cite{zhao.jn:2010}.

\section{Compliance theory}

\begin{figure*}[!t]
\centering
\includegraphics[scale=1.51]{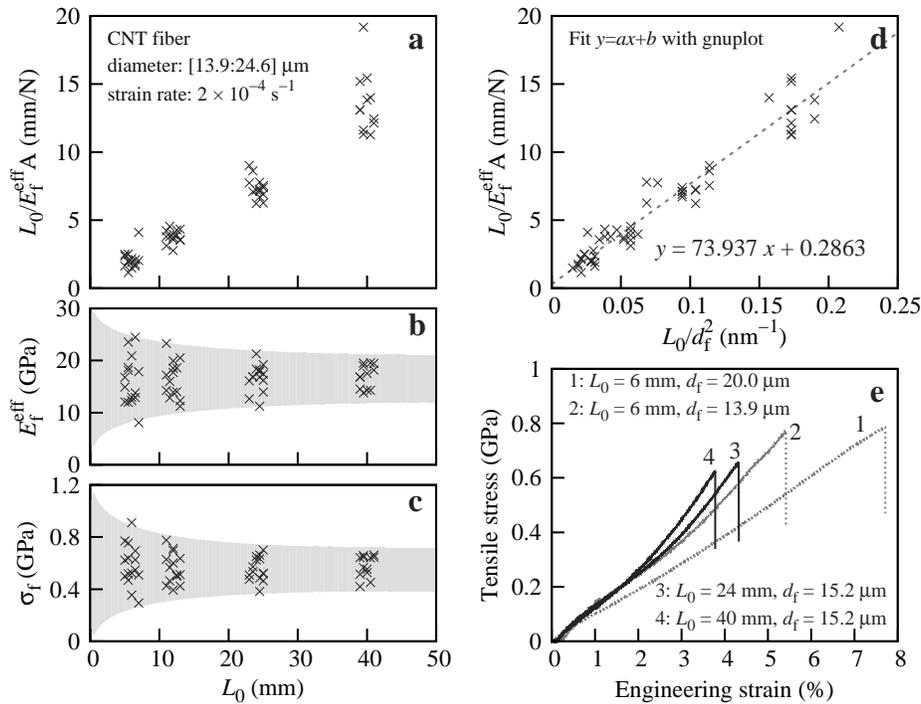}
\caption{Tensile tests of the CNTF-A fibers by the Instron 3365. (a-c) $L_0/\Efeff A$ ($\Ctot$), $\Efeff$, and $\sigf$,
as functions of gauge length $L_0$. For a better show, the data at $L_0$ were plotted scatteredly inside a plot window
[$L_0-1$,$L_0+1$] mm. (d) Compliance evaluation by using the revised fitting method. (e) Typical stress-strain curves
obtained at different gauge lengths.}
\label{fig:instron}
\end{figure*}

Since the engineering strain is not measured directly from the fiber itself but from the crosshead displacement, the
obtained Young's modulus can be viewed as the effective one ($\Efeff$) of two springs in series \cite{kant.m:2014}, as
represented in terms of stiffness by
\begin{equation}
\frac{1}{k_\text{f}^\text{eff}} = \frac{1}{k_\text{sys}} + \frac{1}{k_\text{f}},
\end{equation}
where $k_\text{f}^\text{eff}$, $k_\text{sys}$, and $k_\text{f}$ are the stiffness measured, system stiffness (outside
the sample), and the sample (fiber) stiffness. Compliance is the inverse of stiffness, thus we have
\begin{equation}
\Ctot = \Csys + \Cf.
\end{equation}
According to Hooke's law, we have
\begin{equation}
\Ef = \frac{F/A}{\Delta L/L_0} = k_\text{f}\frac{L_0}{A} = \frac{L_0}{\Cf A}.
\end{equation}
Here $F$ is the tensile force under an elongation distance $\Delta L$, $L_0$ is the gauge length, and $A$ is the cross
sectional area. Thus, with considering the system compliance, the relationship between the measured modulus and the
sample's modulus reads
\begin{equation}
\label{eqn:eff}
\frac{L_0}{\Efeff A} = \Csys + \frac{L_0}{\Ef A}.
\end{equation}
By changing $L_0$ in a wide range and plot the total compliance as a linear function of $L_0$, the slope and
$y$-intercept are thus $1/\Ef A$ and $\Csys$, respectively.

Further, as suggested by Li \textit{et al.} \cite{li1985improvement}, for a sample with a certain diameter variations,
the determination of $\Csys$ can be improved by plotting $\Ctot$ against $L_0/\df^2$, instead of against $L_0$. This
might provide a better way to evaluate CNT fibers as their diameter can be tuned by the level of densification, twisting
degree, and alignment or waviness of nanotube \cite{zhao.jn:2010, jia.jj:2011}.

\section{Results and discussion}

\subsection{Compliance evaluation for low-modulus fibers}

Low-modulus CNT fibers, the CNTF-A and CNTF-B fibers, were first used to evaluate the system compliance for the Instron
and T150 testing machines. Figure \ref{fig:instron}a-c shows the tensile properties for the CNTF-A fibers as functions
of gauge length, obtained from the Instron 3365. When $\Ctot$ was directly plotted against $L_0$ (Figure
\ref{fig:instron}a), the fitting of a function $y=ax+b$ was found to show a huge standard error. To reduce such error, a
large amount of tests should be included. Fortunately, there was a large variation in $\df$ without using solvent
densification, making the revised plotting method applicable \cite{li1985improvement}, as shown in Figure
\ref{fig:instron}d. The gnuplot fitting \cite{gnuplot} output a linear function of $y=73.937x+0.2863$, corresponding to
$\Csys=0.2863$ \si{\milli\meter\per\newton}. Such value was nonzero and had a measurable affect to the true modulus or
stiffness. For example, for a fiber with $\df=13.9$ \si{\um} and $\Efeff = 16.8$ GPa tested at $L_0= 6$ mm, the true
modulus was $\Ef = \Efeff/(1-\pi\df^2 \Csys \Efeff /4L_0) =19.13$ GPa, about 13.8\% higher than the direct measurement.

\begin{figure*}[!t]
\centering
\includegraphics[scale=1.51]{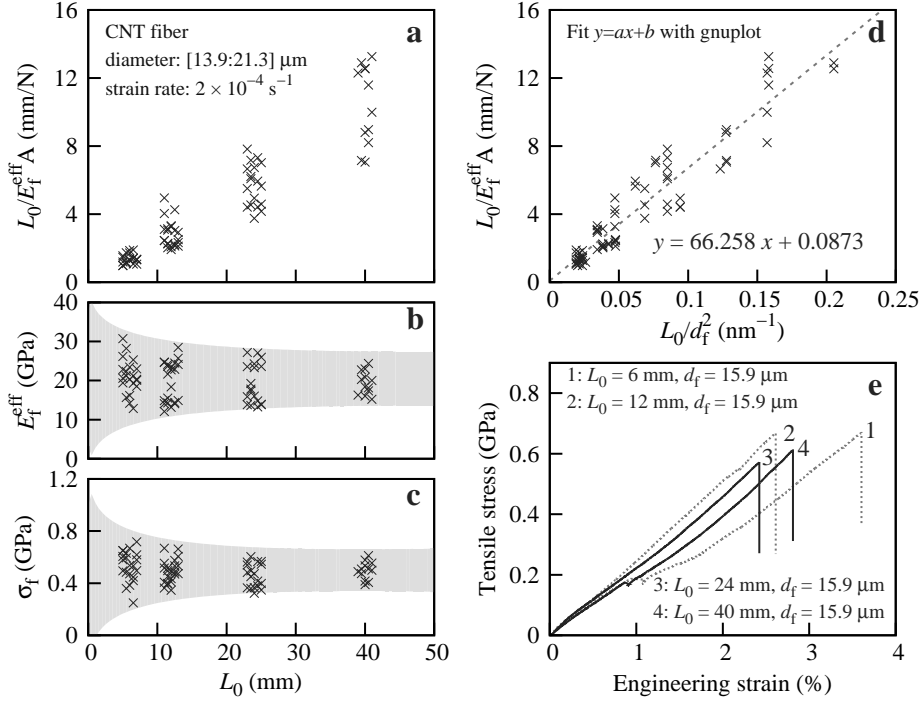}
\caption{Tensile tests and compliance evaluation with the T150 tester. (a-c) $\Ctot$, $\Efeff$, and $\sigf$ as functions
of $L_0$. (d) Compliance evaluation based on the revised plotting method. (e) Four typical stress-strain curves.}
\label{fig:t150}
\end{figure*}

We suspect the way to grip the paper tabs was the reason for the large system compliance, however, not ubiquitous for
all grips. As shown in Figure \ref{fig:tab}c, the pneumatic grip Instron 2712-013 with maximum load of just 5 N could
not hold the paper tab very tightly. Thus under a tensile force of several hundred mN, there could be nonnegligible
deformation at the grip shims, like sliding between the shims, corresponding to a large system compliance. As the effect
of system compliance is more notable for small gauge lengths, the tensile strain is thus often over estimated.
Therefore, the strain at break generally decreased with increasing $L_0$ as shown in Figure \ref{fig:instron}e. Clearly,
for a testing machine with a large system compliance, the gauge length should be even larger, probably over 50 mm.

Different from the pneumatic grips, split-shim grips are used in the T150 testing machine and the paper tabs can be
firmly mounted between the grip shims, see Figure \ref{fig:tab}d. Therefore, the system compliance could be smaller than
that of the Instron tester. Similar to the above evaluation, another low-modulus CNTF-B fiber was tested with the T150,
and Figure \ref{fig:t150} shows the results in the same way as Figure \ref{fig:instron}. The fitting to the dependence
of $\Ctot$ on $L_0/\df^2$ output $y=66.258 x + 0.0873$, corresponding to a small system compliance $\Csys=0.0873$
\si{\milli\meter\per\newton}, just about 1/3 of the Instron's compliance. For a typical test result with $\df=17.7$
\si{\um}, $\Efeff = 21.2$ GPa, and $L_0= 6$ mm, its true modulus was $\Ef = 22.94$ GPa. As the difference was no more
than 8\%, it would not be a severe problem by directly using $\Efeff$ as the true one.

At the same time, the strain at break did not vary greatly at different $L_0$. By considering the sample difference, the
strains at break obtained at $L_0=12$, 24, and 40 mm could all be considered as the true ones, see the stress-strain
curves plotted in Figure \ref{fig:t150}e. Therefore, we suggest a gauge length just larger than 10 mm for the tests on
the T150, much shorter than the suggested value of 50 mm for the Instron's pneumatic grips.

It is notable that in several studies, although quite few yet, a large gauge length of 20--30 mm was used for the tests
with an Instron 5848 tester \cite{zhang.xb_2006, liu.k_20101}.

\subsection{High-modulus fibers}

\begin{figure*}[!t]
\centering
\includegraphics[scale=1.51]{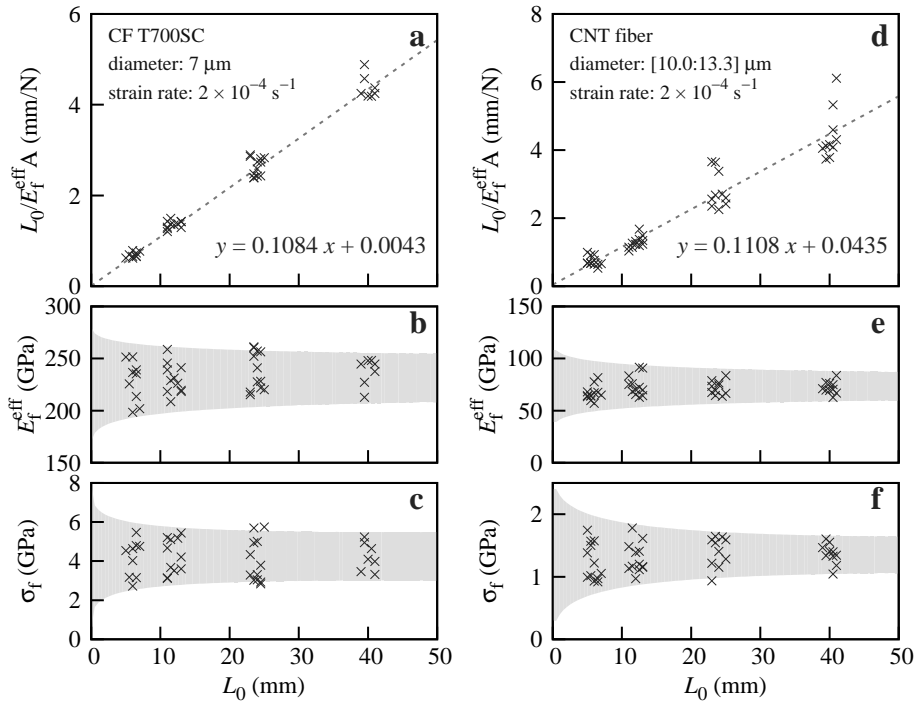}
\caption{$\Ctot$, $\Efeff$, and $\sigf$ as functions of $L_0$ for the CF-T700SC (a-c) and CNTF-C fibers (d-f),
respectively.}
\label{fig:cf}
\end{figure*}

Figure \ref{fig:cf} shows the results for the CF-T700SC and CNTF-C fibers, tested on the T150 tester. According to
Equation \ref{eqn:eff}, the increase in fiber modulus could make the compliance evaluation much easier due to the
decrease in the value of $\Ctot$. Therefore, the fitting was directly based on the dependence of $\Ctot$ on $L_0$. For
the carbon fiber, the T150 system compliance was 0.0043 \si{\milli\meter\per\newton}, and it was just a bit larger,
0.0435 \si{\milli\meter\per\newton} for the CNTF-C fiber. These compliance values were all negligible, indicating that
even a small gauge length could be safe to test the fiber's mechanical property. For a suggestion on the gauge length,
an analysis on the data dispersion is required, as discussed below.

Interestingly, as compared to Figure \ref{fig:t150}, the T150 system compliance decreased with increasing the fiber
modulus. However, this is not a general relationship. The system compliance usually comes from the ``softness'' of the
grips, like the pneumatic grips as an occasional case. Whether such softness can be reflected in the tensile test
depends on the tensile load applied on the fiber. The T150 compliance was nearly zero for the CNTF-B, CNTF-C, and
CF-T700SC fibers because all the failure force was not larger than 250 mN (Table \ref{tab:fiber}), half of the maximum
load of T150, which is hard to induce a measurable deformation either to the rigid grip shims or to the thick photo
paper.

\subsection{Data dispersion}

For the four groups of test described above, there was a common phenomenon that the data dispersion for both $\Efeff$
and $\sigf$ decreased with increasing $L_0$, see Figure \ref{fig:instron}b,c, Figure \ref{fig:t150}b,c, and Figure
\ref{fig:cf}b,c,e,f. For a short gauge length, it became more possible to avoid structural defect along the fiber, and
thus super-high strength or modulus can be occasionally obtained. On the contrary, once the fiber contains structural
defects, there is a considerable chance to grip nearby a defect and thus induce a brittle fracture behavior with an
underestimated tensile strength. The large dispersion of strength data could be an important reason that super-high
strengths were obtained with a very short gauge length of 1--2 mm \cite{koziol.k:2007, wang.jn:2014}.

When the gauge length was sufficiently large, the difference in test samples could be minimized. Thus a reduced
dispersion of strength data was obtained in our tests. From the results of the three groups of CNT fiber, we suggest
again a safe gauge length of at least 10 mm.
 
\subsection{Weibull analysis}

Weibull statistics were used to rank the relative fiber strength versus probability of failure of the fibers to obtain a
measure of the variability in fiber strength \cite{weibull1951, chawla2005}. According to the Weibull analysis, the
probability of survival of a fiber at a stress (fiber strength) $\sigma$ is given by
\begin{equation}
P(\sigma) = \exp \left[ -\left( \frac{\sigma}{\sigma_0}\right)^m\right],
\end{equation}
where $\sigma_0$ is defined as the characteristic strength which corresponds to $P(\sigma_0) = 1/e = 0.37$, and $m$ is
the Weibull modulus. The higher Weibull modulus, the lower the strength dispersion. Ranking of $\sigma$ among $N$
results is performed by using an estimator
\begin{equation}
P(\sigma)_i = 1 - \frac{i}{N+1},
\end{equation}
which describes the probability of survival corresponding to the $i$th strength value. By substituting this estimator
into the previous equation, we get
\begin{equation}
\label{eqn:weibull}
\ln \ln \left[\frac{N+1}{N+1-i}\right] = m \ln (\sigma) - m \ln (\sigma_0).
\end{equation}

\begin{figure}[!t]
\centering
\includegraphics[scale=1.60]{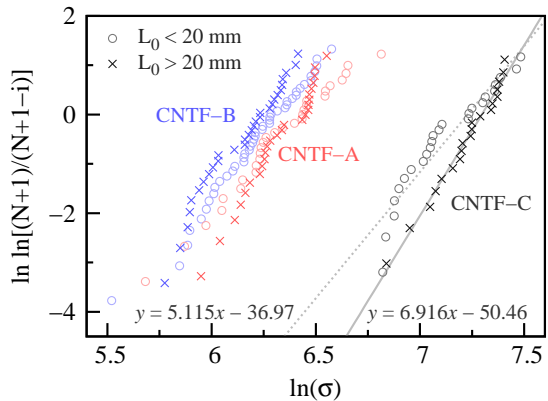}
\caption{Weibull analysis for different CNT fibers. Each fiber was divided into two groups according to the gauge
length. As an example, the linear fittings for the CNTF-C fiber are also plotted to show the effect of gauge length.}
\label{fig:weibull}
\end{figure}

Figure \ref{fig:weibull} shows the Weibull analysis for the three types of CNT fiber according to Equation
\ref{eqn:weibull}, where the data were divided into two groups for each fiber: $L_0 = 6$ and 12 mm and $L_0=24$ and 40
mm, respectively. For each plot, the slope of the linear fitting (by gnuplot \cite{gnuplot}) corresponded to the Weibull
modulus $m$. For example, the CNTF-C fiber exhibited a large Weibull modulus $m=6.916$ when $L_0 > 20 $ mm, and
$m=5.115$ when $L_0 < 20$ mm. (We didn't group the data by a threshold of 10 mm just in order to keep sufficient results
for each group.) This means a large gauge length could lower the strength dispersion. As the $y$-intercept value was a
measurement of $\sigma_0$, we obtained the characteristic strength of $e^{36.97/5.115}=1377$ MPa and $e^{50.46/6.916} =
1475$ MPa for the two groups of CNTF-C fiber. These values were more accurate to their simple averages (1268 and 1383
MPa respectively) as the probability of survival was included.

\section{Conclusion}

The system compliance of testing machine was evaluated for CNT fibers and carbon fibers. For accurate measurement on
fiber modulus, strength, and strain at break, a large heed should be paid to the gripping method and sample's gauge
length. For an ultrafine fiber with a diameter of several to $\sim$20 \si{\um}, the Keysight T150 tester showed great
advantage due to its negligible system compliance. By considering the data dispersion, a safe gauge length of $>$10 mm
was suggested for CNT fibers and carbon fibers, and the Weibull analysis should be used to obtain the statistically
averaged strength.

\bigskip

\begin{acknowledgments}
We thank financial supports from the National Natural Science Foundation of China (11302241, 51561145008, and 21503267)
and the Youth Innovation Promotion Association of the Chinese Academy of Sciences (2015256), and thank Prof.\ Qingwen Li
and Dr.\ Cunyi Xie for their encouragement and suggestive discussion.
\end{acknowledgments}

\end{document}